# Dynamic subgrid-scale LES model for turbulent non-Newtonian flows: *a priori* and *a posteriori* analyses of Burgers turbulence

E. Amani[*], A. Ahmadpour, M.J. Aghajari

*Department of Mechanical Engineering, Amirkabir University of Technology (Tehran Polytechnic), Iran*

**Abstract**

Large Eddy Simulation (LES) of turbulent non-Newtonian flows involves two additional closures, namely the Non-Newtonian SubGrid-Scale (NNSGS) stress tensor and filtered viscosity. Here, dynamic closures are proposed for NNSGS, eliminating the need for model calibration. In addition, for the primary evaluation of LES closures, two canonical case studies are designed by the extension of Newtonian forced Burgers turbulence to include power-law viscosity rheology. The characteristics of the proposed non-Newtonian turbulence are studied carefully using direct numerical simulation. For instance, it is revealed that the shear-thinning effect intensifies the small-scale motions and elevates the energy spectrum function at high wave-numbers. The subsequent *a priori* and *a posteriori* studies manifest that the NNSGS modeling is important for the present test cases. The omission of this term results in the under-prediction of kinetic energy due to the substantial backscatter effect of NNSGS. The proposed dynamic Smagorinsky based closure is recommended, based on the correlation coefficient measure in the *a prio*ri tests and the energy spectrum function in the *a posteriori* tests, for LES of turbulence at high shear-thinning rheology.

**Keywords:** Large eddy simulation (LES); Power-law viscosity; Shear-thinning; Sub-grid scale (SGS) model; non-Newtonian Burgers equation

[*] Corresponding author. Address: Mechanical Engineering Dept., Amirkabir University of Technology (Tehran Polytechnic), 424 Hafez Avenue, Tehran, P.O.Box: 15875-4413, Iran. Tel: +98 21 64543404. Emails: eamani@aut.ac.ir (E. Amani) ali.ahmadpour@aut.ac.ir (A. Ahmadpour), mj_aghajari@aut.ac.ir (M.J. Aghajari).

## 1. Introduction

Turbulent Non-Newtonian Flows (TNNFs) are encountered in many industrial applications, such as heavy crude oil production and transportation, polymer processing, the flow of suspensions in closed conduits, sewages transportation pipelines, drilling mud recirculation systems, food industries, as well as biological applications, like blood flow through arterial stenosis [1]. A large portion of non-Newtonian fluids is classified as pure viscose or inelastic fluids, for which the apparent viscosity is a function of shear rate, the second invariant of the strain-rate tensor. As a result, the broad range of velocity fluctuations and scales present in a turbulent flow brings about viscosity fluctuations over a broadband range. These viscosity fluctuations make the simulation of TNNFs more complicated than the Newtonian counterparts.

The most straightforward and accurate procedure of TNNF simulation is Direct Numerical Simulation (DNS). A substantial body of knowledge on TNNF has been developed over the past two decades with the aid of DNS [2-10], e.g., the shear-thinning effect leads to drag reduction, turbulence transition delay, increased turbulence anisotropy, etc. Recently, Singh et al. [6] pointed out the importance of high-shear rate rheological data for accurate prediction of wall-bounded TNNF due to the existence of the near-wall high-shear rate regions. Despite these improvements in the understanding of TNNF nature, DNS is still deemed impractical for large-scale high-Reynolds-number industrial applications. This emphasizes the need for tractable closures of Reynolds-Averaged-Navier-Stokes (RANS) or Large Eddy Simulation (LES) types.

From early works in the context of RANS models, Wangskarn et al. [11] included non-Newtonian effects in a zero-equation turbulence closure based on Prandtl's mixing length theory for predicting TNNF of thermal entrance region in circular pipes. Malin [12] reported that the standard $k - \varepsilon$ two-equation model, without any changes to account for the non-Newtonian effects, cannot predict friction factor of turbulent pipe flow correctly; therefore, by introducing a new damping function that depended on non-Newtonian material parameters, acceptable results were obtained. Pinho [13] derived second-order moment closure equations for TNNF of a generalized Newtonian fluid, including additional unclosed terms in the momentum equation, i.e., the mean apparent viscosity and non-Newtonian stress tensor associated with the fluctuation of the apparent viscosity, and additional non-Newtonian terms in turbulence transport equations. Then, they proposed a new closure for mean apparent viscosity and presented a simplified $k - \varepsilon$ RANS model by dropping non-Newtonian stress tensor and eliminating non-Newtonian terms from turbulent kinetic energy equation. To include non-Newtonian effects, damping functions, which were modified in their later investigation [14], were suggested.

The weakness of two-equation-based closures, e.g., the ones used in references [11-14], is the inherent inability to describe the turbulence anisotropy. This issue was concerned in subsequent studies [15-17], leading to the development of four-equation turbulence models for TNNFs. In a seminal work, Gavrilov and Rudyak [15] proposed a four-equation $k - \varepsilon - \zeta - f$ model accounting for non-Newtonian stress tensor and non-Newtonian terms in turbulence transport equations. Their model used no damping function and produced excellent results. They used a closure, developed and tested in their previous work [18], for the mean apparent viscosity. Furthermore, in more practical problems, they showed that considering additional non-Newtonian terms in the model equations results in more reliable predictions.

The other type of turbulence models, i.e., LES, is based on filtering and has been proven to be able to capture more details of a turbulent flow in comparison with the RANS approach [19, 20]. Several studies have focused on LES of practical TNNFs, e.g., turbulent blood flow in arterial stenosis [1, 21, 22], power-law turbulent pipe flow [23], viscoelastic FENE-P (finitely extensible nonlinear elastic in the Peterlin approximation) turbulent flow [24, 25]. In contrast to RANS studies, almost all LES of power-law TNNFs used standard Newtonian SubGrid-Scale (SGS) models ignoring the additional closures imposed by apparent viscosity fluctuation, namely the filtered apparent viscosity and non-Newtonian stress tensor. The first study seeking to delve into these topics in the context of LES is probably the work by Otha and Miyashita [26]. They tried to include the non-Newtonian effects in the static Smagorinsky model through damping functions. However, as noted by Gavrilov and Rudyak [15], the main issue of models using local damping functions is that they do not differentiate between the viscous and nonviscous damping effects of the wall, leading to the false activation of damping functions. Therefore, reliable SGS modeling for LES of power-law fluid TNNF is still a challenge.

The first aim of the present work is to develop a closure for non-Newtonian SGS stress tensor. In the context of LES, damping functions were gradually replaced with dynamic models that do not require the calibration of model coefficients. Thus, the idea is to propose a dynamic closure for the non-Newtonian SGS stress tensor for the first time. It is standard practice to assess new SGS models through canonical case studies. Burgers equation and Burgers turbulence (Burgelence) is one of the most popular test cases due to its similarity to flow turbulence, e.g., nonlinear structure, energy spectrum, intermittency of energy dissipation, etc. The free and forced Burgers equations have been extensively used for evaluating SGS models, e.g., [27-29]. All previous efforts considered the Burgers equation with Newtonian-like condition, i.e., with constant viscosity. The second aim of the current study is to propose a non-Newtonian variant of the forced Burgers equation, which is useful for evaluating additional closures arisen in TNNF.

The rest of the paper is organized as follows. In section 2, governing equations required for LES of TNNF are presented, and different closures, including the new dynamic SGS non-Newtonian stress tensor, are detailed. In section 3, the non-Newtonian variants of the Burgers equation and their selected parameters are proposed. In section 4, the numerical algorithm and the simplifications required to apply the models to the 1D Burgers equation are put forward. In section 5, after the validation and verification of the numerical solver, *a priori* and *a posteriori* analyses of different SGS closures are reported, and the characteristics of the models are studied carefully. Finally, the main conclusions of the present study are summarized in section 6.

## 2. Mathematical modelling

The incompressible power-law non-Newtonian fluid flow in the absence of body forces is governed by conservation laws for mass and linear momentum as follows:

$$\frac{\partial u_j}{\partial x_j} = 0 \tag{1}$$

$$\frac{\partial u_i}{\partial t} + \frac{\partial}{\partial x_j}(u_j u_i) = -\frac{\partial p}{\partial x_i} + \frac{\partial \tau_{ij}}{\partial x_j} \tag{2}$$

where, $u_i$ and $p$ are the velocity vector and (kinematic) pressure, respectively, and $\tau_{ij}$ is the extra (kinematic) stress tensor which is a function of the shear rate, $S$, according to the power-law non-Newtonian constitutive relation as:

$$\tau_{ij} = 2\nu S_{ij} \tag{3}$$

$$\nu = K_\nu S^{n-1} \tag{4}$$

Here, $S_{ij}$, $\nu$, $K_\nu$, and $n$ are strain-rate tensor, apparent (kinematic) viscosity, consistency factor, and power-law index, respectively, and

$$S_{ij} = \frac{1}{2}\left(\frac{\partial u_i}{\partial x_j} + \frac{\partial u_j}{\partial x_i}\right); \quad S = (2S_{ij}S_{ij})^{\frac{1}{2}} \tag{5}$$

Note that the summation convention is applied to all repeated indices in the present formulation and the "kinematic" properties are obtained by dividing the original ones by the density. For an incompressible flow, it is convenient to express the equations in terms of kinematic pressure, kinematic stress, and kinematic viscosity. Hereafter, the word kinematic is implied for these properties. LES formulation is obtained by filtering Eqs. (1) and (2), which results in the following LES equations, with the approximation of homogenous filter:

$$\frac{\partial \bar{u}_j}{\partial x_j} = 0 \qquad (6)$$

$$\frac{\partial \bar{u}_i}{\partial t} + \frac{\partial}{\partial x_j}(\bar{u}_i \bar{u}_j) = -\frac{\partial \bar{p}}{\partial x_i} + \frac{\partial \tau_{ij}^{\bar{v}}}{\partial x_j} + \frac{\partial(-\tau_{ij}^r)}{\partial x_j} + \frac{\partial \tau_{ij}^N}{\partial x_j} \qquad (7)$$

where $\overline{(\ )}$ indicates the filtered quantities, and $\tau_{ij}^{\bar{v}}$, $\tau_{ij}^r$, and $\tau_{ij}^N$ are the Newtonian-like filtered-viscosity (FV) stress, residual or SGS stress, and Non-Newtonian SGS (NNSGS) or Non-Newtonian Residual (NNR) stress tensors, respectively, which are defined by:

$$\tau_{ij}^{\bar{v}} = 2\bar{v}\bar{S}_{ij} \qquad (8)$$

$$\tau_{ij}^r = \tau_{ij}^R - \frac{1}{3}\tau_{kk}^R \delta_{ij}; \quad \tau_{ij}^R = \overline{u_i u_j} - \bar{u}_i \bar{u}_j \qquad (9)$$

$$\tau_{ij}^N = 2\overline{vS_{ij}} - 2\bar{v}\,\bar{S}_{ij} \qquad (10)$$

$$\bar{S}_{ij} = \frac{1}{2}\left(\frac{\partial \bar{u}_i}{\partial x_j} + \frac{\partial \bar{u}_j}{\partial x_i}\right); \quad S = (2\bar{S}_{ij}\bar{S}_{ij})^{\frac{1}{2}} \qquad (11)$$

Therefore, the filtered non-Newtonian formulation imposes two additional unknowns, namely the filtered apparent viscosity, $\bar{v}$, (see Eq. (8)) and NNSGS stress, $\tau_{ij}^N$, in addition to the conventional SGS stress, $\tau_{ij}^r$, for a turbulent Newtonian fluid flow. Therefore, turbulent non-Newtonian fluid flows involve two additional closure problems. For a turbulent Newtonian fluid flow of constant viscosity, $\bar{v} = v$ and $\tau_{ij}^N$ vanishes. In all previous LES of TNNF, e.g., [22, 23, 26], $\tau_{ij}^N$ has been neglected and $\overline{v(S)} = v(\bar{S})$ simply used. In the present work, models are proposed for these additional unknowns and evaluated with *a priori* and *a posteriori* studies using DNS results. The closure of different terms is detailed in the following subsections.

*2.1. SGS modeling*

The effect of unresolved small turbulence scales should be accounted for through the modeled SGS stress tensor in LES. Due to the advantage of dynamic SGS models, here, the dynamic Smagorinsky model [30] is adopted as:

$$-\tau^r_{ij} = 2\nu_r \bar{S}_{ij} \qquad (12)$$

where $\nu_r$ is the eddy viscosity of residual motions which is given by:

$$\nu_r = C_{DS} \bar{\Delta}^2 \bar{S} \qquad (13)$$

and $\bar{\Delta}$ is the length scale (filter width) corresponding to the primary (or main) filtering operation and the dynamic model coefficient, $C_{DS}$, is computed from [31, 32]:

$$C_{DS} = \frac{\langle l_{ij} M_{ij} \rangle}{\langle M_{mn} M_{mn} \rangle} \qquad (14)$$

where the tensors $l_{ij}$ and $M_{ij}$ are defined by:

$$l_{ij} = \widetilde{\bar{u}_i \bar{u}_j} - \tilde{\bar{u}}_i \tilde{\bar{u}}_j \qquad (15)$$

$$M_{ij} = 2\bar{\Delta}^2 \widetilde{|\bar{S}| \bar{S}_{ij}} - 2\tilde{\bar{\Delta}}^2 |\tilde{\bar{S}}| \tilde{\bar{S}}_{ij} \qquad (16)$$

and $\widetilde{\overline{(.)}}$ indicates the double filter operation with the filter width of $\tilde{\bar{\Delta}}$ or the application of the secondary filter $\widetilde{(.)}$ with the filter width of $\tilde{\Delta}$ on the primary-filtered quantities and $\langle . \rangle$ operator refers to the spatial averaging over the homogenous direction(s). Here, similar to the choices used in many flow solvers, like ANSYS FLUENT (www.ansys.com), OpenFOAM (www.OpenFoam.org), etc., the primary filter is taken as the grid (implicit) filter, the secondary filter is chosen with the width twice the primary filter $\tilde{\Delta}= 2\Delta$, and $\tilde{\tilde{\Delta}}$ value required in Eq. (16) is chosen as $\tilde{\tilde{\Delta}}= \tilde{\Delta}$. Note this last assumption imposes no difficulty in the dynamic model formulation since the difference of $\tilde{\tilde{\Delta}}/\Delta$ and $\tilde{\Delta}/\Delta$ is absorbed in the dynamic model constant, $C_{DS}$.

*2.2. Filtered apparent viscosity modeling*

The filtered apparent viscosity, $\bar{\nu}$, is required for the calculation of $\tau_{ij}^{\bar{\nu}}$, based on Eq. (8), and for NNSGS stress closure described in the next section. According to Eq. (3), the filtered viscosity is written as:

$$\bar{\nu} = K_\nu \overline{S^{2\frac{n-1}{2}}} \tag{17}$$

A simplification like the one used for RANS models by Gavrilov and Rudyak [15] is considered here:

$$\overline{S^{2\frac{n-1}{2}}} \approx \bar{S}^{2\frac{n-1}{2}} \tag{18}$$

Now, $\overline{S^2}$ can be decomposed into two parts using Eq. (5):

$$\overline{S^2} = 2\overline{S_{ij}S_{ij}} = 2\bar{S}_{ij}\bar{S}_{ij} + 2(\overline{S_{ij}S_{ij}} - \bar{S}_{ij}\bar{S}_{ij}) = \bar{S}^2 + 2(\overline{S_{ij}S_{ij}} - \bar{S}_{ij}\bar{S}_{ij}) \tag{19}$$

The second term on the right-hand side is unresolved and can be approximated by $\varepsilon_r/\bar{\nu}$, where $\varepsilon_r$ is the SGS dissipation rate, therefore, $\overline{S^2} \approx \bar{S}^2 + \varepsilon_r/\bar{\nu}$. Under the local equilibrium assumption, $\varepsilon_r \sim P_r = |-\tau_{ij}^r \bar{S}_{ij}|$, and the closure is finally achieved by:

$$\bar{\nu} = K_\nu \overline{S^2}^{\frac{n-1}{2}} \tag{20}$$

$$\overline{S^2} = \bar{S}^2 + \frac{\varepsilon_r}{\bar{\nu}} \tag{21}$$

$$\varepsilon_r = |-\tau_{ij}^r \bar{S}_{ij}| \tag{22}$$

Eqs. (20) and (21) are a system of two non-linear equations and two unknowns, $\overline{S^2}$ and $\bar{\nu}$. Eliminating $\overline{S^2}$, the filtered apparent viscosity can be computed from:

$$\left(\frac{1}{K_\nu}\right)^{\frac{2}{n-1}} \bar{\nu}^{\frac{n+1}{n-1}} - \bar{S}^2 \bar{\nu} = \varepsilon_r \tag{23}$$

Eq. (23) is called Non-Linear Filtered Viscosity (NLFV) model hereafter. This approach is similar to the one used by Gavrilov and Rudyak [15] in the context of RANS formulation with the difference that they finally dropped the first term on the right-hand side of Eq. (21) assuming the negligible contribution of this term in RANS. By this assumption, $\bar{\nu}$ can be explicitly obtained in terms of the strain-rate tensor. However, in the context of LES, we found out that the first term of Eq. (21) is important and should be kept. This results in Eq. (23) for $\bar{\nu}$ which is implicit and should be solved using a root-finding approach, e.g., the Newton-Raphson method.

The second approximation, called Linear Filtered Viscosity (LFV) model, examined here is to neglect the second term on the right-hand side of Eq. (21) that leads to the following closure:

$$\bar{\nu} \approx K_\nu (\bar{S})^{n-1} \tag{24}$$

Note that Eq.(24) is the same as the assumption $\overline{\nu(S)} = \nu(\bar{S})$ in Eq. (4) used in all previous LES studies.

*2.3. NNSGS modeling*

The main contribution of the present study is developing an NNSGS closure. To utilize the advantages of dynamic closure and maintain the consistency with our choice of SGS model in section 2.1, a dynamic-Smagorinsky-like closure is developed for $\tau_{ij}^N$ in this section. Using the eddy-viscosity assumption:

$$\tau_{ij}^N = 2\nu_N \bar{S}_{ij} \tag{25}$$

Here, $\nu_N$ is the eddy-viscosity of non-Newtonian unresolved motion. Its value can be positive or negative, which introduces the possibility of backscatter in the model. In contrast to the SGS eddy viscosity, $\nu_r$, it will be seen that this parameter is predicted, by the dynamic approach, to be more frequently negative than positive for a shear-thinning fluid. This finding is consistent with the presumption used by Gavrilov and Rudyak [15] in the context of a four-equation RANS approach. They presumed that $\nu_N$ is proportional to $(n-1)$ which makes the value of $\nu_N$ to be negative for shear-thinning fluids ($n < 1$). Using a simple dimensional analysis, $\nu_N$ can be described in terms of $\bar{\Delta}$ and $\bar{S}$ as:

$$\nu_N = C_N \bar{\Delta}^2 \bar{S} \tag{26}$$

where $C_N$ is the model coefficient determined by a dynamic-Smagorinsky-like approach here. Using the secondary filter operation $\widetilde{(.)}$ and NNSGS of double filter $\widetilde{\overline{(.)}}$, i.e., $T_{ij} = 2\widetilde{\overline{\nu S_{ij}}} - 2\widetilde{\overline{\nu}} \widetilde{\overline{S}}_{ij}$, the resolved non-Newtonian stress is defined as:

$$l_{ij}^N = T_{ij} - \widetilde{\tau_{ij}^N} = 2\widetilde{\overline{\nu} \overline{S}_{ij}} - 2\widetilde{\overline{\nu}}\, \widetilde{\overline{S}}_{ij} \tag{27}$$

Then by the least square error minimization like the one used for the dynamic Smagorinsky coefficient [32], it can be shown that the coefficient $C_N$ is governed by:

$$C_N = -\frac{\langle l_{ij}^N M_{ij} \rangle}{\langle M_{mn} M_{mn} \rangle} \tag{28}$$

where, $M_{ij}$ is given by Eq. (16).

Here, another dynamic model is assessed, which has been inspired by the dynamic Bardina model formulation [31, 33]:

$$\tau_{ij}^N = L_{ij}^{0,N} + \tau_{ij}^{k,N} \tag{29}$$

where, $\tau_{ij}^{k,N}$ is unresolved part of NNSGS tensor and $L_{ij}^{0,N}$ is presented with the analogy to Leonard stress tensor as:

$$L_{ij}^{0,N} = 2\overline{\bar{\nu}\bar{S}_{\iota J}} - 2\overline{\bar{\nu}}\bar{\bar{S}}_{ij} \tag{30}$$

which is resolved. The unresolved part of NNSGS tensor is then modeled by:

$$\tau_{ij}^{k,N} = 2C_{NB}\bar{\Delta}^2 \bar{S}\,\bar{S}_{ij} \tag{31}$$

where, the Bardina coefficient, $C_{NB}$, is determined by Eq.(28) replacing $l_{ij}^N$ with $l_{ij}^{BN}$:

$$l_{ij}^{BN} = l_{ij}^N - H_{ij}^N \tag{32}$$

$$H_{ij}^N = 2\left(\widetilde{\overline{\bar{\nu}\bar{S}_{\iota J}}} - \widetilde{\overline{\bar{\bar{\nu}}}}\,\widetilde{\overline{\bar{\bar{S}}}}_{ij} - \left(\widetilde{\overline{\nu S_{\iota J}}} - \widetilde{\bar{\nu}}\widetilde{\bar{S}_{\iota J}}\right)\right) \tag{33}$$

## 3. Non-Newtonian Stochastic Burgers turbulence

Burgers equation and Burgers turbulence is a well-known test case to evaluate new discretization schemes and turbulence models. Here, the stochastic or random-force-driven Burgers equation proposed by Chekhlov and Yakhot [34] is extended to a power-law variable viscosity condition. The one-dimensional Non-Newtonian Stochastic Burgers Equation (NNSBE), which is the simplified form of Eqs. (2)-(4), dropping the pressure gradient term and adding a random force term, $f(x,t)$, can be written as follow:

$$\frac{\partial u}{\partial t} + u\frac{\partial u}{\partial x} = \frac{\partial \tau}{\partial x} + f(x,t); \quad 0 \leq x \leq 2\pi \tag{34}$$

$$\tau = \tau_{11} = 2\nu S_{11} = 2\nu \frac{\partial u}{\partial x} \tag{35}$$

$$\nu = K_\nu(S)^{n-1}; \quad S = (2S_{11}S_{11})^{\frac{1}{2}} = \sqrt{2}\left|\frac{\partial u}{\partial x}\right| \tag{36}$$

where, the periodic boundary condition is assumed at $x = 0, 2\pi$ and the initial condition is $u(x,t) = 0$ and the random forcing function $f$ is given by [29, 34]:

$$f(x,t) = 2\sqrt{\frac{D_0}{\Delta t}}\mathcal{F}^{-1}\left\{|k|^{\frac{\beta}{2}}\hat{f}(k)\right\} \tag{37}$$

where $\mathcal{F}^{-1}\{.\}$ represents the inverse Fourier transform and $D_0$, $\Delta t$, $\beta$ and $\hat{f}(k)$ are the noise amplitude, sampling time step, noise spectral slope, and the $k$th Fourier coefficient of the white noise [29]. Based on the recommendation by Basu [29], the model parameters are chosen as $D_0 = 1 \times 10^{-6}$ and $\beta = -3/4$. Our *a priori* analysis of the LES models is performed through DNS of NNSBE, i.e., Eqs. (34)-(37). The non-Newtonian model parameters are carefully chosen so as to produce a certain level of viscosity fluctuations due to turbulence. Here, $K_\nu = 3 \times 10^{-5}$ and the shear-thinning effect is studied in the range of $n = 0.2 - 1.0$. To avoid unphysically large or small values of viscosity, the power-law viscosity is constrained by the minimum value of $\nu_{min} = 5 \times 10^{-6}$ and the maximum value of $\nu_{max} = 10^{-3}$.

For *a posteriori* analysis, filtering Eqs. (34)-(37) and introducing the closures used in section 2, the LES formulation of NNSBE can be written as:

$$\frac{\partial \bar{u}}{\partial t} + \frac{\partial}{\partial x}(0.5\bar{u}\bar{u}) = \frac{\partial \tau^{\bar{\nu}}}{\partial x} + \frac{\partial(-\tau^r)}{\partial x} + \frac{\partial \tau^N}{\partial x} + \bar{f} \tag{38}$$

$$\tau^{\bar{\nu}} = 2\bar{\nu}\bar{S}_{11} = 2\bar{\nu}\frac{\partial \bar{u}}{\partial x}; \quad -\tau^r = 2\nu_r\frac{\partial \bar{u}}{\partial x}; \quad \tau^N = 2\nu_N\frac{\partial \bar{u}}{\partial x} \tag{39}$$

Hereafter, the second term on the left-hand side of Eq. (38) is called Advection (Adv.) and the terms on the right-hand side of Eq. (38) are called the Filtered-Viscosity Stress Gradient (FVSG),

Residual Stress Gradient (RSG), Non-Newtonian Residual Stress Gradient (NNRSG), and Filtered Force (FF) terms, respectively. The only difference with the 3D formulation, reported in section 2, is that in 1D SGS stress, $\tau^r$, is defined as $\tau^r = 0.5(\overline{uu} - \bar{u}\bar{u})$ instead of the 3D SGS stress, Eq. (9), simplified to 1D, i.e., $(\overline{uu} - \bar{u}\bar{u})$, owing to the presence of coefficient 0.5 in the advection term of 1D NNSBE, Eq. (38). This leads to the definition of the resolved stress as $l_{11} = 0.5(\widetilde{\bar{u}\bar{u}} - \tilde{\bar{u}}\tilde{\bar{u}})$ instead of Eq. (15) simplified to 1D, i.e., $(\widetilde{\bar{u}\bar{u}} - \tilde{\bar{u}}\tilde{\bar{u}})$, and $\varepsilon_r = |-2\tau^r \bar{S}_{11}|$ instead of Eq. (22) simplified to 1D, i.e., $|-\tau^r \bar{S}_{11}|$. All other closures and relations introduced in sections 2.1 to 2.3 can be directly simplified to the filtered 1D NNSBE, Eq. (38), substituting $\bar{S}_{ij} = \bar{S}_{11}$.

To check the model performance in another relevant test case, the second test case of forced non-Newtonian Burgers turbulence is defined by choosing $f(x,t) = 0$ in Eq. (34) and implementing a forcing function in spectral (Fourier) space. For this purpose, the solution $u$ (or $\bar{u}$ in the case of LES) is transformed to the spectral space by $\hat{u}(k,t) = \mathcal{F}\{u(x,t)\}$, where $\mathcal{F}\{.\}$ indicates the Fourier transform of a function and $\hat{u}(k,t)$ is the kth Fourier coefficient of $u(x,t)$, and injecting energy into several low-wave-number modes to mimic the turbulent energy production at large turbulence scales. Here, the forcing (energy injection) is conducted by keeping the magnitude of 4 Fourier coefficients equal to one at every iteration, leaving their phase unchanged, as:

$$|\hat{u}(k,t)| = 1; \quad k = 3, 4, 5, \text{ and } 6 \tag{40}$$

where $|.|$ returns the magnitude of a complex variable. For this case, $K_v = 0.2$, $\nu_{\min} = 5 \times 10^{-3}$, and $\nu_{\max} = 0.38$ are chosen and $n$ is studied in the range $0.4 - 1.0$. This test case is called Non-Newtonian Spectral Forcing Burgers Equation (NNSFBE), here.

## 4. Numerical method

DNS, i.e., the solution of Eqs. (34)-(36), and LES, i.e., the solution of Eqs. (38)-(39), of non-Newtonian Burgelence, are obtained for the *a priori* and *a posteriori* tests, respectively. Accurate assessment of the theoretical models requires keeping the computational errors at a low level. For this aim, all spatial derivatives in all *a priori* and *a posteriori* studies are calculated by the ($n^{th}$-order) spectral method [35]. For the *a priori* studies, the derivatives of filtered properties are computed in the LES grid. The temporal derivatives are discretized by the second-order Adams-Bashforth scheme [36]. The aliasing error in calculating products (non-linear terms) is also removed by the pseudo-spectral padding algorithm [37]. For DNS of the first test case, i.e., NNSBE e, the grid resolution of $N_{DNS} = 8192$ nodes and time step of $dt = 10^{-4}$ is required to resolve all scales. For the *a posteriori* test of this case, LES is performed on a uniform grid with 1024 nodes using implicit (grid) primary filtering procedure, i.e., $dx_{LES} = 8dx_{DNS}$. Therefore, the *a priori* test is conducted using a box filter of width $\Delta = 8dx_{DNS}$ on the DNS results to be consistent with the *a posteriori* study. For the DNS of the second test case, the grid resolution of $N_{DNS} = 2048$ nodes and time step of $dt = 1.25 \times 10^{-6}$ is used. For the *a priori* tests of this case, a box filter with the width-to-grid-size ratio of $\Delta/dx_{DNS} = 8$ is incorporated similar to the first case. All *a posteriori* LES solutions are obtained on $dx_{LES} = 8dx_{DNS}$ using the implicit primary filter.

It should be noted that the anti-aliasing implemented by the pseudo-spectral method is even more crucial for the non-Newtonian SBE, compared to Newtonian SBE studied in the previous works, due to the appearance of the second nonlinear term, i.e., the nonlinear diffusion term the first term on the right-hand side of Eq. (34). To highlight this fact, the energy spectrum function calculated from DNS of NNSBE with $n = 0.2$, with and without anti-aliasing of the diffusion term, is shown in Figure 1. As can be seen from this figure, the level of energy especially at scales close to the grid scale or Nyquist scale (the highest wave-number shown in the diagram) is under-predicted without anti-aliasing of the diffusion term. This can be explained by noting that without diffusion term anti-aliasing, the level of diffusion near the Nyquist scale unphysically increases due to aliasing by the scales smaller than the grid scale. This increased diffusion dampens the velocity fluctuations at resolved scales close to the Nyquist scale and results in the observed under-prediction of energy at the high-wavenumber part of the spectrum.

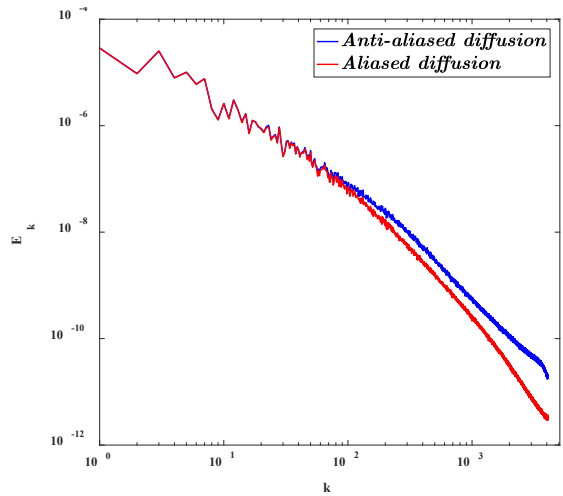

Figure 1 DNS of NNSBE: The energy spectrum function versus wavenumber for $n = 0.2$. The spectrum is time-averaged over $50 < t < 100$.

For LES, by default, the clipping is considered for modeling SGS term, i.e., $C_{DS} \geq 0$, unless stated otherwise. Note that this is not to be implemented on the dynamic coefficient of NNSGS stress, i.e., $C_N$, because, NNRSG can have both diffusion- or anti-diffusion-like (backscatter) behaviors.

## 5. Results and discussion

For the validation of the developed solver, a limit case of the Burgers equation, Eqs. (34) and (35), without forcing function and with constant viscosity of $\nu = 0.5$, decaying from the initial condition of $u(x, 0) = 10\sin(x)$, is considered. This limit case has an analytical solution obtained by Benton and Platzman [38]. The numerical solution obtained by the present spectral solver is compared with the analytical one in Figure 2. The fine agreement between the results advocates the validation of the present solver.

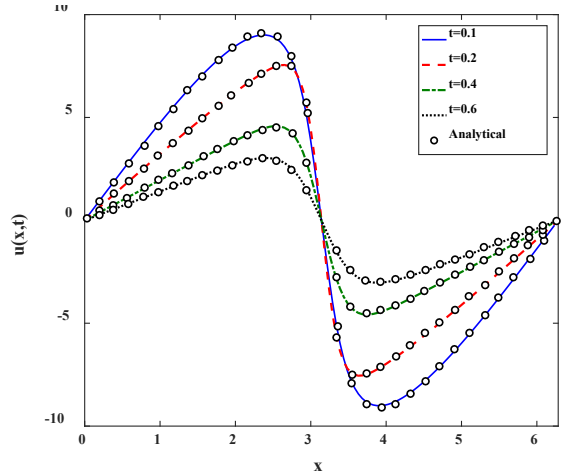

Figure 2 Validation: The solution of Burgers equation for a limit case ($v = 0.5$, $f = 0$, and $u(x, 0) = 10\sin(x)$) at different times. The comparison of the present computation (lines) and the analytical solution [38] (symbols).

As another case for code verification, the DNS of stochastic forced Burgers equation under a Newtonian-like condition, i.e., Eqs. (34)-(37) with constant viscosity of $v = 0.5 \times 10^{-5}$, is chosen (i.e., the limit condition of $n = 1$ and $K_v = 0.5 \times 10^{-5}$ set in the solver). For this Newtonian-like test case, a numerical solution with a similar spectral solver has been reported by Basu [29] on a uniform 8192-cell grid. The numerical solution obtained by the present computation is compared with the reference solution by Basu in Figure 3. The reported energy spectrum is obtained by time-averaging the results during $50 < t < 200$ exactly in the same procedure as detailed by Basu [29]. The agreement between these solutions corroborates the validity of the present numerical computation.

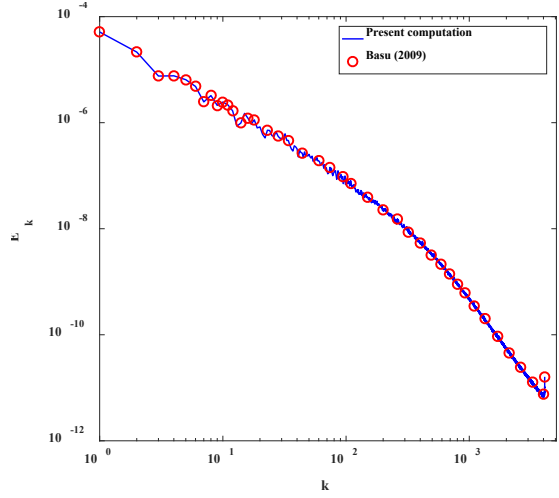

Figure 3 Verification: The energy spectrum of DNS of Newtonian stochastic forced Burgers equation. The comparison of the present computation and the reference numerical solution [29].

*5.1. DNS of NNSBE and NNSFBE*

In this section, first, the results of DNS of the newly proposed NNSBE test case are presented and analyzed. The energy spectrum of different cases with different values of the power-law index, $n$, obtained from DNS is plotted in Figure 4. Here, the spectrum is computed by:

$$E_k(k) = 0.5 \langle \hat{u}'(k,t)\hat{u}'^*(k,t) \rangle_T \qquad (41)$$

where $\hat{u}'(k,t)$ is the $k^{th}$ Fourier coefficient of the fluctuation of the solution around its mean value, i.e., $u' = u - \langle u \rangle_T$, at time $t$, the Asterisk superscript indicates the complex conjugate, and $\langle . \rangle_T$ stands for the time-averaging operator. For the spectrum presented here, time-averaging is performed over the time interval of $50 < t < 100$, sampled every 0.1 time unit. As observed in Figure 4, with increasing the shear-thinning effect (the decrease of $n$) the level of energy, especially at low wavenumbers, grows. This indicates that the shear-thinning effect amplifies the small-scale motions and results in stronger velocity fluctuations (total energy) in the NNSBE test case. This can be attributed to the escalation of viscosity fluctuations by decreasing $n$, which is illustrated in Figure 5(a). A similar trend has also been reported for fluid flow equations in the previous studies where the rise of viscosity Root-Mean-Square (RMS) by increasing the shear-thinning effect triggers the increase of stream-wise velocity RMS [9, 26]. The energy surplus at lower $n$ values for the NNSBE is concentrated at higher wavenumbers (small-scale motions) as can be seen in Figure 4. In Figure 5(b), the NNSGS stress of the filtered solution, computed *a priori* from DNS results using a box filter of width $\Delta/dx = 8$, is reported for different power-law indices. As can be seen from the results, the level of this stress augments with the intensification of the shear-thinning effect, suggesting the need for an SGS model to account for the NNSGS stress term as viscosity variations elevate.

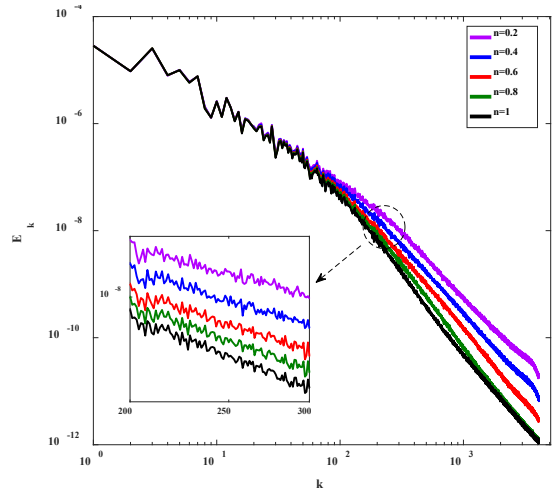

Figure 4 DNS of NNSBE: The energy spectrum function versus wavenumber for different values of power-law index, $n$. The spectrum is time-averaged over $50 < t < 100$.

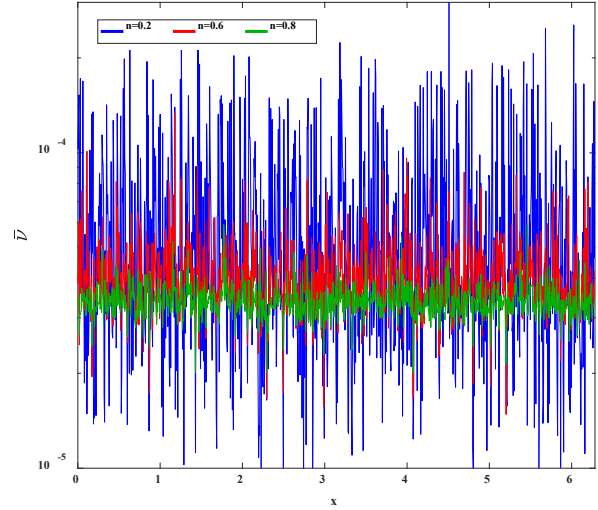

**(a)**

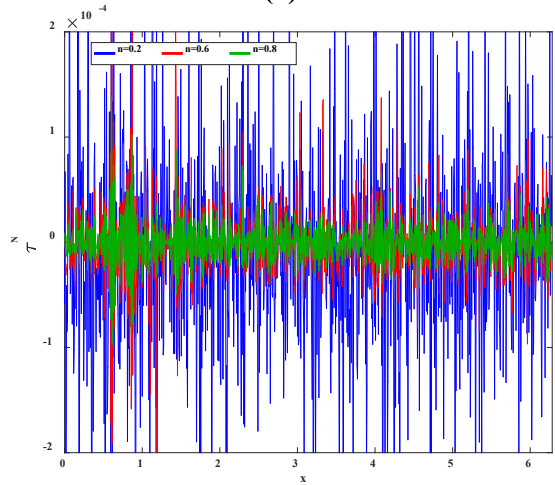

**(b)**

Figure 5 DNS of NNSBE: The filtered viscosity (a) and NNSGS stress (b) versus $x$ at $t = 100$ for different values of power-law index, $n$.

Next, the DNS of the NNSFBE test case is performed for Newtonian and shear-thinning conditions. The energy spectrums of different cases with different values of the power-law index are reported in Figure 6 in conjunction with the spatial variation of viscosity at an instance of simulation in Figure 7. For this case, the energy spectrum is time-averaged over the interval $5 < t < 20$, sampled every 0.1 time unit. Only, values of $n$ larger than 0.6 are considered since the present spectrum is saturated for $n < 0.6$ and much larger grid resolutions would be required to perform DNS of these cases due to the distribution of a considerable amount of energy over scales smaller than the present DNS grid scale. We ignored such simulations due to the large computational cost of the NNSFBE solution compared to NNSBE. Nevertheless, the cases considered here are sufficient to draw note-worthy conclusions. For this case, similar to the NNSBE test case, the kinetic energy of fluctuations especially at smaller scales (larger wave-numbers) grows with decreasing $n$, however, the level of fluctuation intensity growth is much larger than the NNSBE test case and the change of shear-thinning rheology has a much larger impact on the solution. In both test cases, the RMS of viscosity fluctuation increases with decreasing $n$ which results in the observed growth of kinetic energy; however, for the NNSBE test case the mean viscosity slightly grows while for the NNSFBE test case mean viscosity reduces with decreasing $n$. This indicates that the variation of mean viscosity by changing the rheology of the system depends on the type of forcing function employed. Therefore, the introduced two cases with different forcing natures can pose different challenges for LES models, and it would be worthwhile to assess models using both test cases, i.e., NNSBE and NNSFBE, introduced here.

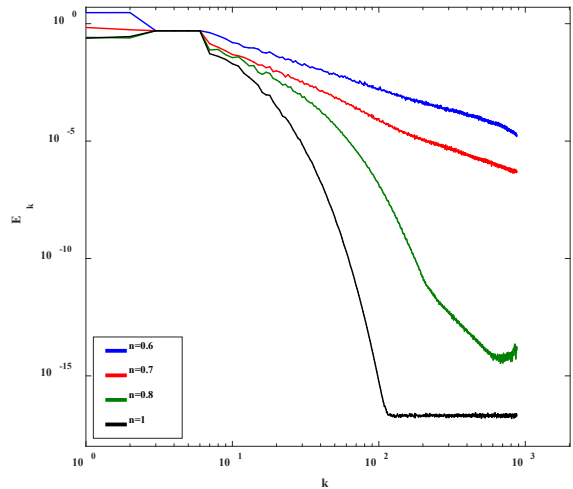

Figure 6 DNS of NNSFBE for different values of power-law index, $n$. The energy spectrum function versus wavenumber. The spectrum is time-averaged over $5 < t < 20$.

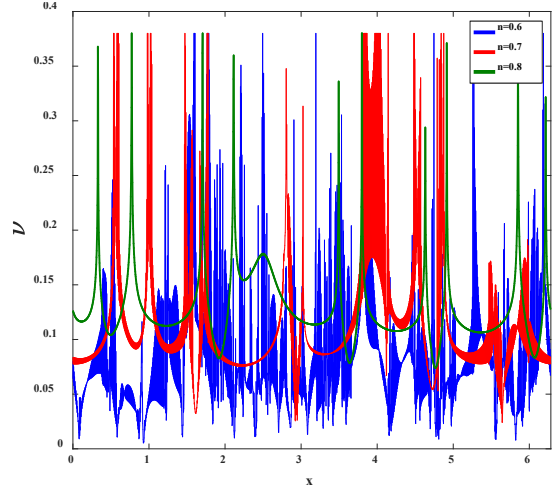

Figure 7 DNS of NNSFBE for different values of the power-law index, $n$. The viscosity versus $x$ at $t = 20$.

*5.2. A priori studies*

In this section, the *a priori* analysis of LES, i.e., Eqs. (38) and (39), of the NNSBE and NNSFBE with the newly proposed dynamic SGS models is performed by post-processing the DNS results obtained in the previous section. For the filtering operation, a box filter with the width-to-grid-size ratio of $\Delta/dx_{DNS} = 8$ is considered. Other numerical details of the *a priori* tests were reported in section 4.

For evaluating the performance of different models of different unclosed terms, i.e., Eq. (39), in the LES equation, the Correlation Coefficient (CC) measure is usually adopted. For the *a priori* study, this measure for a parameter $A$ is defined by:

$$CC(A) = \frac{\langle (A_E - \langle A_E \rangle_{x,T})(A_M - \langle A_M \rangle_{x,T}) \rangle_{x,T}}{\langle (A_E - \langle A_E \rangle_{x,T})^2 \rangle_{x,T}^{1/2} \langle (A_M - \langle A_M \rangle_{x,T})^2 \rangle_{x,T}^{1/2}} \qquad (42)$$

where, the averaging operator $\langle . \rangle_{x,T}$ uses all samples at all space locations $x$ and sampled every 0.1 time unit over the time interval $50 < t < 100$ for the NNSBE case and over $5 < t < 20$ for NNSFBE.

The subscripts "E" and "M" stand for the exact value of the parameter obtained from DNS results and the modeled value of the parameter obtained by applying a specific model on DNS results, respectively. The CC of two models, i.e., dynamic Smagorinsky and dynamic Bardina, for SGS stress, $\tau^r$, is reported in Table 1. According to this table, for the first test case (NNSBE), the correlation coefficient of the dynamic Smagorinsky model is relatively high (close to 0.8) regardless of the value of $n$ index which shows that this model can be regarded as an accurate model (at least based on the CC measure) even for shear-thinning rheological behavior of Burgers equation. The correlation coefficients of the scale similarity dynamic Bardina model are also high (around 0.73) but slightly lower than the ones reported for the dynamic Smagorinsky model. The results of the second case (NNSFBE) show the same trend except at the strong shear-thinning condition ($n = 0.6$). However, due to the saturation of the DNS energy spectrum at $n = 0.6$ (see Figure 6a), the LES of this case with the 8 times coarser grid size is far from being a well-resolved LES. In other words, the low CC at this condition can be regarded as the consequence of a too-coarse grid resolution and grid size inefficacy for a high-quality well-resolved LES rather than the deficiency of the SGS closure.

Table 1 The correlation coefficient (for $\tau^r$) of two SGS models, i.e., dynamic Smagorinsky (Smag.) and dynamic Bardina (Bard.), reported for two case studies, i.e., NNSBE and NNSFBE, and different power-law indices, $n$.

| $n$ | Case: NNSBE | | $n$ | Case: NNSFBE | |
|---|---|---|---|---|---|
| | Smag. | Bard. | | Smag. | Bard. |
| **0.2** | 0.756 | 0.725 | | | |
| **0.4** | 0.792 | 0.744 | **0.6** | 0.386 | 0.561 |
| **0.6** | 0.794 | 0.738 | **0.7** | 0.710 | 0.702 |
| **0.8** | 0.782 | 0.732 | **0.8** | 0.892 | 0.866 |
| **1.0** | 0.772 | 0.741 | **1.0** | 0.922 | 0.997 |

The second closure required for LES of a non-Newtonian Burgers equation is modeling the filtered viscosity, $\bar{\nu}$, contributing to $\tau^{\bar{\nu}}$. The correlation coefficients of the LFV and NLFV models for $\bar{\nu}$ are reported in Table 2. Several conclusions can be made from this data. First, the correlation coefficients of $\bar{\nu}$ closures are much smaller than the ones of $\tau^r$ reported in Table 1. This highlights the importance of efforts to improve the accuracy of modeling this term while the majority of the previous works adopted the simplest possible choice, i.e., LFV. Second, NLFV has a slightly larger CC. However, the level of improvement seems insignificant. To check whether the input of the NLFV model, i.e., $\tau^r$ in $\varepsilon_r = |-2\tau^r \bar{S}_{11}|$ of Eq. (23), rather than the model itself brings about the reported low CCs here, we compared the value of CC for NLFV based on the values of $\tau^r$ obtained by exact relation ($\tau^r = 0.5(\overline{uu} - \bar{u}\bar{u})$) from the DNS results and modeled values by the dynamic Smagorinsky and Bardina closures in Table 2. As it can be observed, the values of CC computed with different aforementioned approaches do not differ significantly, advocating the deficiency of the NLFV model but its inputs. Gavriov and Rudyak [18] reported much noticeable enhancement using a similar closure in the RANS framework. Therefore, the present results may suggest the need for more improvement in $\bar{\nu}$ modeling at least for the LES framework. The conclusion drawn in our *a priori* study is revisited in the *a posteriori* study in the next section. Third, the accuracy of the models decreases by increasing the shear-thinning effect (decreasing $n$). Fourth, the levels of CC are much different for both test cases, NNSBE and NNSFBE. CC is much larger for NNSFBE. Since these two cases are different in the forcing mechanism and turbulent energy production, this can suggest that the accuracy of apparent filtered viscosity closure is problem-dependent and depends on the energy production mechanism or equivalently wall boundaries and geometrical features for fluid flow problems.

Table 2 The correlation coefficient of the NLFV (non-linear filtered viscosity) and LFV (linear filtered viscosity) models for filtered viscosity, reported for two case studies, i.e., NNSBE and NNSFBE, and different power-law indices, $n$. NLFV is considered with three different estimations for $\tau^r$, i.e form DNS, dynamic Smagorinsky, and dynamic Bardina models.

| | **Case: NNSBE** | | | |
|---|---|---|---|---|
| | **NLFV** | | | |
| $n$ | $\tau^r$ (DNS) | $\tau^r$ (Smag.) | $\tau^r$ (Bard.) | **LFV** |
| **0.2** | 0.090 | 0.092 | 0.093 | 0.091 |
| **0.4** | 0.096 | 0.097 | 0.098 | 0.095 |
| **0.6** | 0.137 | 0.138 | 0.139 | 0.130 |
| **0.8** | 0.280 | 0.282 | 0.284 | 0.268 |
| | **Case: NNSFBE** | | | |
| | **NLFV** | | | |
| $n$ | $\tau^r$ (DNS) | $\tau^r$ (Smag.) | $\tau^r$ (Bard.) | **LFV** |
| **0.6** | 0.344 | 0.341 | 0.351 | 0.337 |
| **0.7** | 0.557 | 0.571 | 0.571 | 0.564 |
| **0.8** | 0.916 | 0.919 | 0.919 | 0.918 |

The correlation coefficients of modeled NNSGS stress, $\tau^N$, are reported in Table 3. The CCs of $\tau^N$ are generally smaller than the ones of $\tau^r$ (see Table 1) with the present dynamic models. This highlights the complexity of $\tau^N$ and questions the validity of eddy-viscosity models for its closure. However, the level of this stress is usually much smaller than $\tau^r$ in a fluid flow problem and a more complicated closure for $\tau^N$ may be unnecessary. In other words, the present dynamic model can be a trade-off between cost and accuracy for modeling $\tau^N$ in more practical problems. The dynamic Smagorinsky model generally performs better than the dynamic Bardina model for $\tau^N$, especially at stronger shear-thinning conditions. The other important conclusion is that the accuracy of the dynamic Smagorinsky model is highest at intermediate shear-thinning conditions ($n \sim 0.6 - 0.7$). By increasing $n$ close to the Newtonian condition, the CC of this model decreases slightly. Note, however, the value of $\tau^N$ itself approaches zero as $n$ tends to 1.

Table 3 The correlation coefficient (for NNSGS stress, $\tau^N$) of two NNSGS models, i.e., dynamic Smagorinsky (Smag.) and dynamic Bardina (Bard.), reported for two case studies, i.e., NNSBE and NNSFBE, and different power-law indices, $n$.

| $n$ | Case: NNSBE | | $n$ | Case: NNSFBE | |
|---|---|---|---|---|---|
| | Smag. | Bard. | | Smag. | Bard. |
| 0.2 | 0.292 | 0.144 | | | |
| 0.4 | 0.385 | 0.244 | 0.6 | 0.555 | 0.416 |
| 0.6 | 0.536 | 0.404 | 0.7 | 0.694 | 0.605 |
| 0.8 | 0.524 | 0.486 | 0.8 | 0.454 | 0.822 |

*5.3. A posteriori studies*

The conclusions made in *a priori* studies are not necessarily valid for an LES. Therefore, it is necessary to revisit the findings by *a posteriori* studies. The present *a posteriori* study is performed through an LES using an implicit box filter with $dx_{LES}/dx_{DNS} = 8$ to be comparable with the filter used in the *a priori* test. Other features of the *a posteriori* test were detailed in section 4. To show and compare the importance of different terms in Eq. (38) and their potential to influence the solution, the L2-norm of different terms, $A$, of Eq. (38) is computed by:

$$L2 - norm(A) = \sqrt{\langle A^2 \rangle_{x,T}} \qquad (43)$$

where, the averaging operator $\langle . \rangle_{x,T}$ is the same as the one used in Eq. (42). The L2-norms are plotted against the power-law index in Figure 8 for the NNSBE case. As it can be seen, for the Newtonian case ($n = 1$), the advection (Adv.) is roughly 2.5 times the viscous diffusion (FVSG), and NNRSG is zero as expected. By decreasing $n$ and increasing the shear-thinning effect, keeping the forcing function unchanged, all terms and phenomena involved in the Burgers transport equation intensify. The rate of NNRSG rise with decreasing $n$ is higher than the other terms and this term reaches the level of RSG at $n = 0.2$. The parameters of the present NNSBE have been chosen in such a way that all terms have important contributions to the solution and cannot be neglected. This is essential to evaluate the model performance for different unclosed terms, i.e.,, RSG, FVSG, and NNRSG, in *a posteriori* tests.

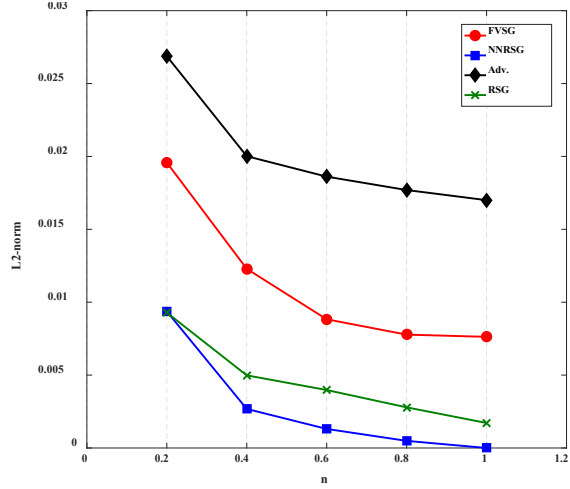

Figure 8 The L2-norm of different terms of LES equation versus the power-law index, $n$, for NNSBE. The closures are the dynamic Smagorinsky model for $\tau^r$ and $\tau^N$, and NLFV for $\tau^{\bar{\nu}}$. Adv.: advection, FVSG: filtered-viscosity stress gradient, RSG: residual stress gradient, and NNRSG: non-Newtonian residual stress gradient.

LESs of the NNSBE case are performed for different power-law index values and different modeling closures, and the energy spectrums of the solutions are compared against the spectrum of the Filtered DNS solution (FDNS) in Figure 9. For all LESs in this figure, the SGS stress is closed using the dynamic Smagorinsky model. The effects of filtered viscosity and NNSGS modeling are investigated in this figure. For examining the effect of filtered viscosity ($\bar{\nu}$) modeling, the NNSGS stress is modeled using the dynamic Smagorinsky model while $\bar{\nu}$ is accounted for by LFV, NLFV, and NLFV with backscatter (BS) models. To include the effect of backscatter in the NLFV model, the input of NLFV, i.e., $\tau^r$ in $\varepsilon_r = |-2\tau^r \bar{S}_{11}|$ of Eq. (23), is computed without clipping $C_{DS}$. Note that this is done only for the calculation of $\bar{\nu}$ by NLFV and RSG term is still incorporated using the standard clipped $\tau^r$. Based on the comparison reported in Figure 9, the NLFV model slightly improves the predictions, especially at strong shear-thinning rheological conditions (smaller $n$ values). This supports the conclusion obtained in the *a priori* analysis based on a different measure, i.e., the correlation coefficient. The NLFV (BS) slightly improves the predictions of NLFV as the shear-thinning effect grows, however, this model suffers from much higher computational time and lower robustness. This model was not converged for the strongest shear-thinning condition considered in this study ($n = 0.2$). Therefore, the standard NLFV model can be considered as the best option and a robust alternative to its BS variant.

The other conclusion that can be made from the data provided in Figure 9 is that the NNSGS stress ($\tau^N$) modeling is important for the present test case. As can be observed in this figure, the solutions neglecting this term ($\tau^N = 0$) results in the under-prediction of the energy spectrum especially when the shear-thinning effect grows. This is attributed to the anti-diffusion or backscatter effect of the NNSGS term which elevates the level fluctuations and kinetic energy. Therefore, its omission leads to a smaller energy level and under-prediction of the energy spectrum especially at the scales close to the filter width, larger frequencies in the spectrum. To corroborate the anti-diffusion effect of NNSGS, the average non-Newtonian eddy-viscosity, $\langle \nu_N \rangle_{x,T}$, for different cases with different power-law indices is reported in Figure 10. The averaging operator $\langle . \rangle_{x,T}$ is the same as the one used in Eq. (42). This figure indicates that the value of $\nu_N$ is negative on the average sense and supports the anti-diffusion-dominated effect of NNSGS. Besides, this anti-diffusion effect strengthens as the shear-thinning behavior grows. Therefore, the inclusion of an NNSGS model is necessary for the present test case and the proposed dynamic models improve the predictions.

In Figure 9, to explore the effect of NNSGS closure type, two different models, including dynamic Smagorinsky and dynamic Bardina models, are incorporated for $\tau^N$. Based on the results reported in Figure 9, the dynamic Smagorinsky model generally performs better than the dynamic Bardina.

The dynamic Bardina model shows large over-predictions of the energy spectrum as $n$ reduces. Therefore, the dynamic Smagorinsky NNSGS model can be proposed as the better option which is consistent with the conclusion drawn from the *a priori* analysis in section 5.2.

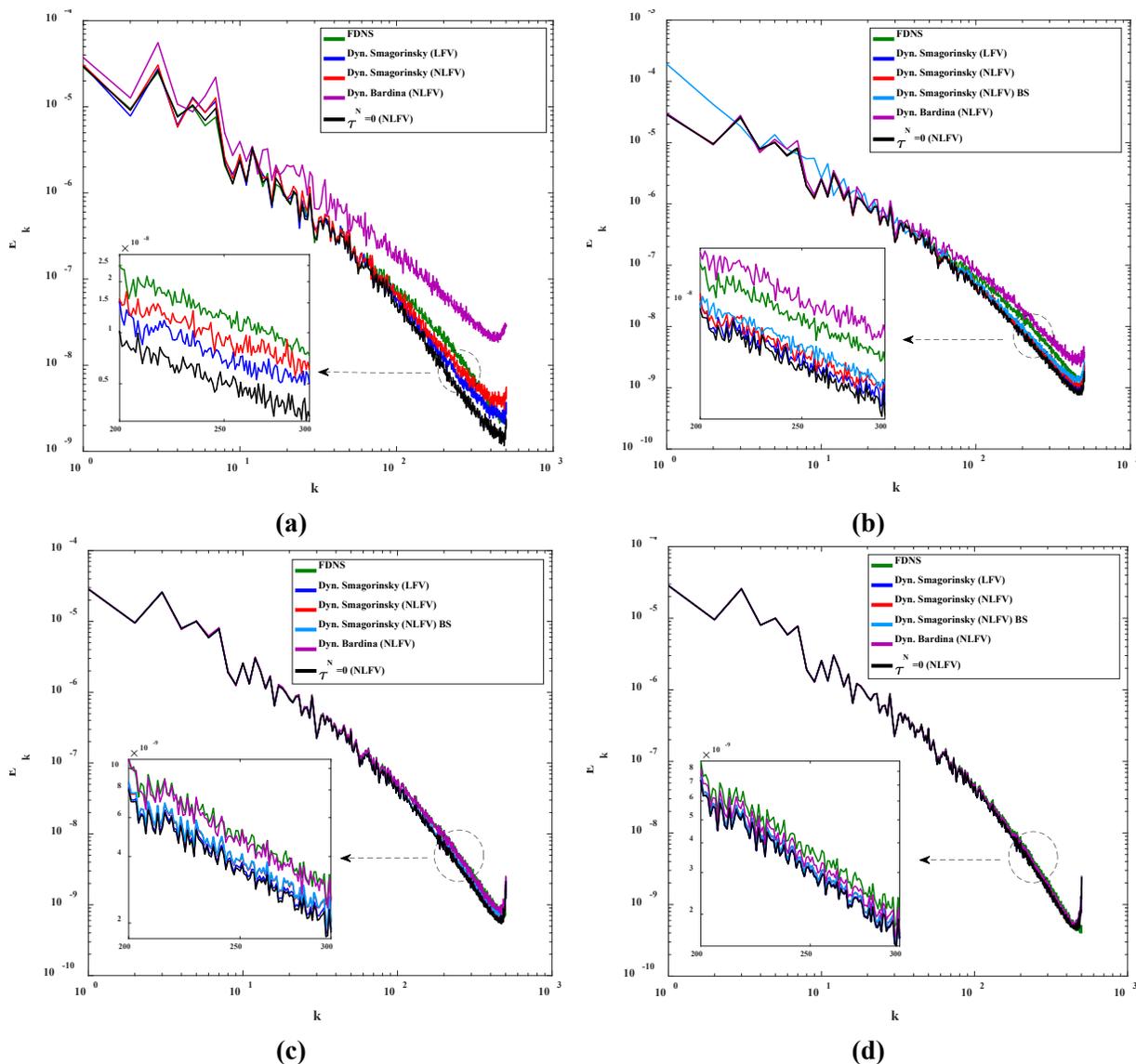

Figure 9 LES of NNSBE: The energy spectrum for a) $n = 0.2$, b) $n = 0.4$, c) $n = 0.6$, and d) $n = 0.8$. The spectrum is time-averaged over $50 < t < 100$. The comparison of FDNS and LES with the dynamic Smagorinsky SGS model. The effect of filtered viscosity modeling, i.e., LFV, NLFV, and NLFV with backscatter (BS), and NNSGS modeling, i.e., dynamic Smagorinsky, dynamic Bardina, and no model ($\tau^N = 0$). For $n = 0.2$, NLFV (BS) was not converged.

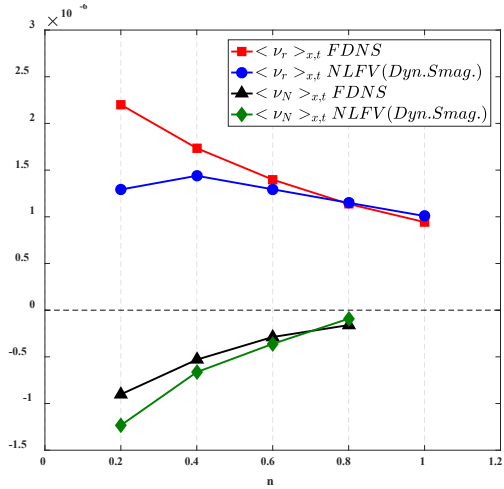

Figure 10 The average SGS and NNSGS viscosities versus the power-law index, $n$, in LES of NNSBE. The comparison of DNS and LES (with NLFV-dynamic-Smagorinsky closure).

The energy spectrum functions obtained from LES of the second test case, i.e., NNSFBE, are plotted against the FDNS in Figure 11. These results support the conclusions made in the case of the former test case. The examination of the LES closures for the second database, emphasizes the importance of the NNSGS stress ($\tau^N$) modeling. The omission of this term leads to the under-prediction of the energy spectrum. The anti-diffusion impact of the dynamic Bardina model is too high at high shear-thinning conditions and results in the over-prediction of the energy spectrum, and the dynamic Smagorinsky closure for the NNSGS stress would be the best choice.

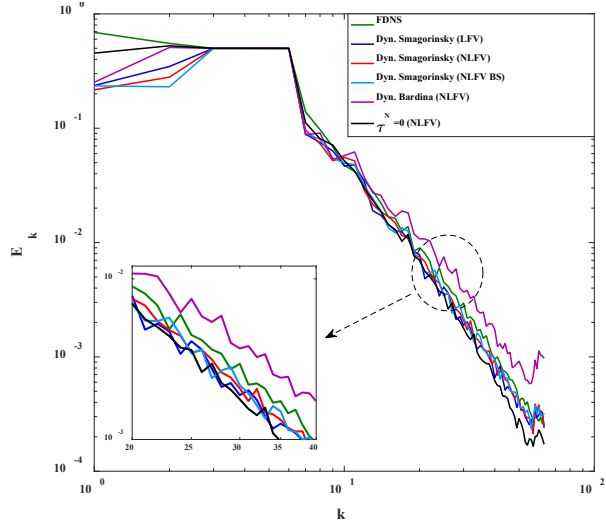

Figure 11 LES of NNSFBE: The energy spectrum for $n = 0.7$. The spectrum is time-averaged over $5 < t < 20$. The legend description is the same as the one in Figure 9.

## 6. Conclusion

In the present study, for the LES of non-Newtonian turbulent flows, dynamic closures were proposed for the NNSGS stress tensor. Besides, a non-linear model for filtered viscosity was extended to LES from similar models in RANS. To assess the models, two canonical case studies were designed by including the power-law-viscosity rheological behavior in Burgers turbulence generated by two different forcing mechanisms. The direct numerical simulation of non-Newtonian Burglenece, for the first time, revealed that the shear-thinning effect increases the kinetic energy of turbulence by amplifying the small-scale (high-frequency) part of the energy spectrum. In addition, keeping the forcing function unchanged and increasing the shear-thinning effect, all advection, filtered viscosity diffusion, SGS stress, and NNSGS stress budgets elevate. Next, *a priori* and *a posteriori* analyses showed that the NNSGS stress term is important for the present test cases, and its omission brings about the under-prediction of the kinetic energy. This is because the anti-diffusion-dominated effect of NNSGS tends to raise the velocity fluctuation intensity. The dynamic Bardina model over-predicted this effect at strong shear-thinning conditions. The proposed dynamic Smagorinsky based closure was the best choice, among the models evaluated, and resulted in the best correlation coefficient in the *a priori* tests and energy spectrum function in the *a posteriori* tests. This model is recommended for LES of turbulent non-Newtonian flows at high shear-thinning circumstances in future studies. The results also showed that the extended non-linear model for the filtered viscosity does not offer significant improvement over the simple widely-used linear closure, in the LES framework, and suggested the need for more effort into improved filtered viscosity models.

**Formatting of funding sources**

This research did not receive any specific grant from funding agencies in the public, commercial, or not-for-profit sectors.